# D.V. Vlasov, L.A. Apresyan, V.I. Krystob, T.V. Vlasova

# Investigation of conductivity switching upon action of monoaxial pressure on plasticized PVC films

**Summary**

The effect of conductivity switching of wideband polymers -plasticized PVC films under the influence of mono axial pressure is experimentally investigated. For various plasticizers the value of monoaxial pressure, causing jumps of conductivity on four and more orders, changes from units to hundreds bars, and the effect remains at a thickness of films of an order of hundreds micron, that is on two orders more than critical thickness for others wideband polymers. In addition to the reasons stated earlier on the interpretation of anomalies of plastic compounds conductivity, the phenomenological electron-molecular model of dynamic traps is considered, in which local transfer of charges is carried out by mobile segments of the plasticized polymer molecules.

1. Введение

В предшествующих работах авторов [1-5] было показано, что в антистатических полимерных пластифицированных ПВХ-пленках во внешнем электрическом поле возникают спонтанные и индуцированные переходы в состояние с относительно высокой электропроводностью (СВП), превышающей исходную проводимостью на четыре и более порядков. В отличие от большинства опубликованных на эту тему результатов такие переходы в СВП реализуются не только на тонких (порядка и менее микрона), но и на более толстых пленках, толщиной порядка 30-100 мкм. Исследование перехода в СВП в относительно толстых полимерных пленках, по мнению авторов, способно внести определенную ясность в плане гипотез, используемых для создания моделей СВП, поскольку оно снижает возможное влияние состояния электродов, что позволяет сконцентрировать внимание на аномалиях и специфике проводимости самой полимерной пленки. Хотя к настоящему времени в литературе рассмотрено несколько моделей переключения проводимости в полимерных пленках (см., напр.,обзоры [6,7]) , начиная от простейшей модели замыкания электродов или прорастания металлических усов – дендритов и кончая предложенной в [8] моделью, рассматривающей собственно аномалии проводимости полимера, а именно формирования проводящих каналов из метастабильных экситонов (электронно-дырочных пар, см.также недавние работы [9,10]), механизм возникновения СВП в полимерных пленках, как под воздействием одноосного давления, так и под действием поля, по общему мнению [6,7] нельзя считать окончательно установленным.

Ранее в [1-5] авторами было экспериментально найдено, что в области антистатической проводимости для ПВХ пластикатов не имеют места ограничения, связанные с измеренной для других широкозонных полимеров малой и обычно не превышающей 2-3 мкм критической толщиной [11,12], выше которой переходы в СВП не наблюдаются. Отсутствие этого ограничения для пластикатов, по мнению авторов, может оказаться еще более существенным для исследования переходов полимерных пленок в СВП под действием одноосного давления [6-9]. Так, при исследовании воздействия одноосного давления на тонкие пленки (толщиной десятки и сотни нм) разделить влияние эффектов, связанных с аномалиями проводимости полимера и артефактов типа замыкания шероховатых, наклонных или неплоских поверхностей или образования микротрещин и дендритов представляется весьма проблематичным.

В соответствии с вышесказанным, в данной работе исследуются закономерности перехода в СВП относительно «толстых» пленок ПВХ - пластикатов при наложении одноосного давления. Полученные результаты позволили подтвердить и развить предложенную нами ранее в [1,2] качественную модель, предложив новый электронно-молекулярный механизм локального переноса заряда для интерпретации электропроводности пластифицированных ПВХ материалов.

## 2. Экспериментальная часть

Ниже изложены результаты экспериментальных исследований зависимости электропроводности от величины одноосного давления, выполненные на относительно толстых пленках ПВХ пластикатов от 30 до 700 и более микрон. В работе использованы гомогенные пленки пластикатов, полученные методом полива из раствора ПВХ на плоские металлические и стеклянные поверхности, что достаточно подробно описано в работах [1-5].

В экспериментах одноосное давление создавалось лабораторным винтовым прессом, причем образец пленки пластиката фиксировался в одном из нескольких вариантов стандартной кольцевой ячейки, предназначенной для проведения гостированных измерений проводимости полимерных пленок [1-5]. Ячейка с площадью измерительного электрода 4.7 см$^2$ помещалась в винтовой пресс с устройством измерения силы F, откалиброванным в пределах от 0 до 1500 Н и приложенной симметрично к центру плоской поверхности верхнего электрода, что обеспечивало создание одноосного давления в направлении, перпендикулярном плоскости электродов ячейки. В кольцевой ячейке для создания одноосного давления использовались только внутренние электроды, предназначенные для измерения объемного сопротивления, в то время как внешний кольцевой электрод перемещался свободно, т.е. использовался в качестве экрана и фактически не участвовал в создании давления и в измерениях. Электрические измерения тока и проводимости образцов пластикатов выполнялись на программно задаваемых импульсных последовательностях зондирующего напряжения с периодом порядка единиц секунд и в интервале напряжений от +10В до -10В. Постоянно приложенный потенциал можно было задавать как частный случай импульсной последовательности. Использование импульсных зондирующих последовательностей позволяло определить характер релаксационных и нелинейных искажений отклика испытуемого образца. Применение более высоких частот подачи импульсов ограничивалось характерными временами RC цепочек схемы. Блок схема электрической части экспериментальной установки приведена на Рис.1 . Для упрощения компьютер, осуществлявший сбор данных и перепрограммирование генератора импульсных последовательностей, на схеме не показан.

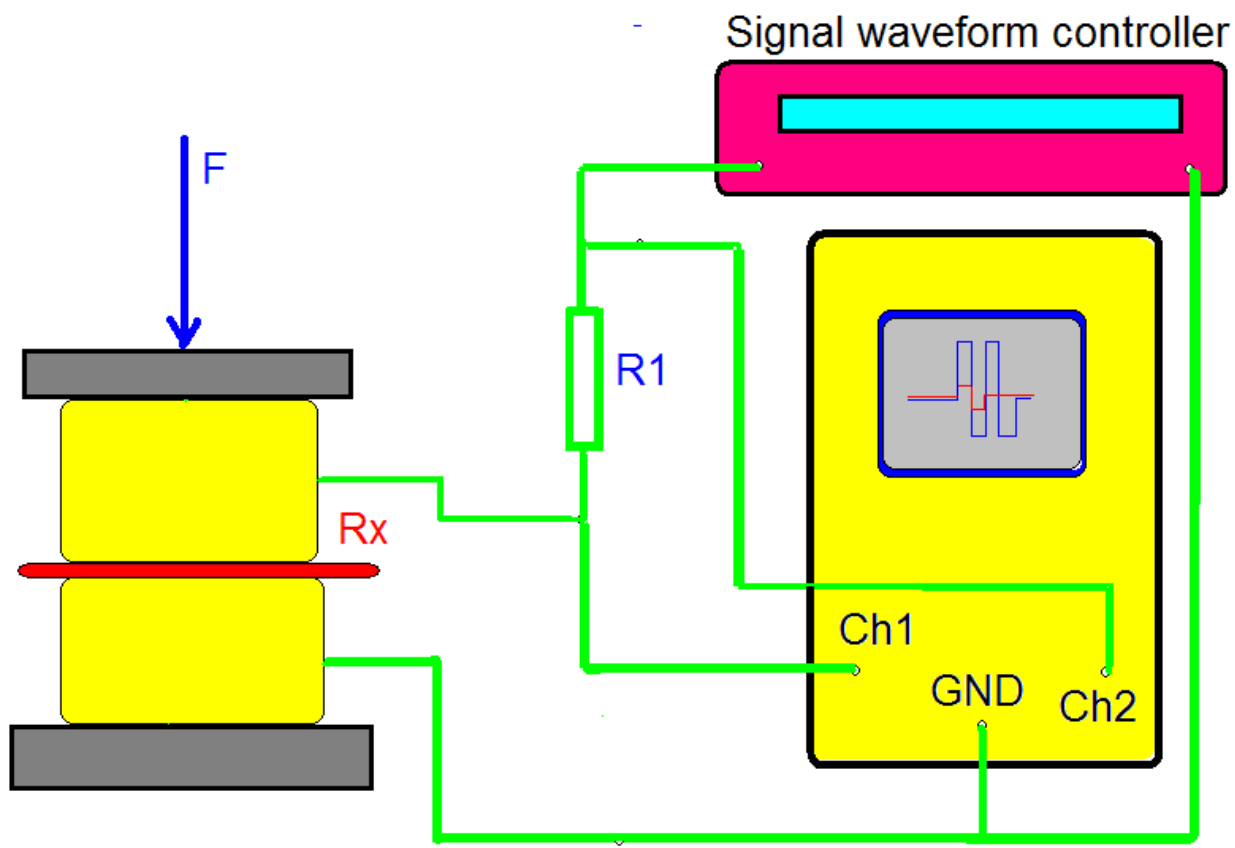

Рис 1. Схема электрической части экспериментальной установки для измерения проводимости образцов пленок пластикатов ПВХ при воздействии одноосного давления.

Выбор балластного сопротивления осуществлялся, как правило, таким образом, чтобы в отсутствии давления все подаваемое напряжение «садилось» на образце и осциллограммы на обоих каналах совпадали. Характерная осциллограмма, соответствующая начальному этапу измерений, приведена на Рис.2.

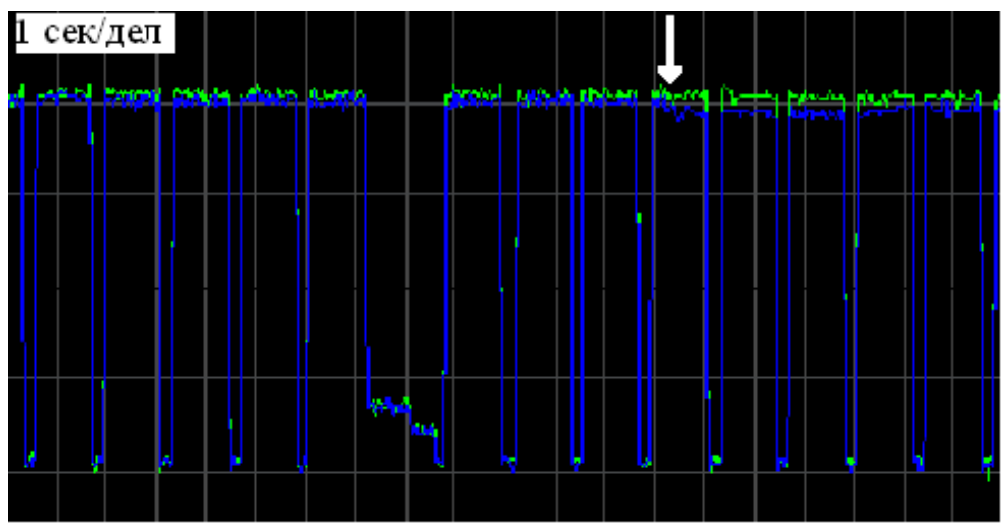

Рис.2. Характерная осциллограмма (развертка по времени 1 сек/дел) напряжений на измерительной установке. В начале развертки при «нулевом» давлении все падение напряжения происходит на образце (полимерной пленке), причем в отмеченной стрелкой точке развертки при приложении силы порядка одного Ньютона сопротивление образца уменьшается и различие амплитуд импульсов становится заметным.

Небольшое шумовое различие сигналов осциллограмм в отсутствие приложенной силы сопряжено с наличием балластного сопротивления R1<<Rx (реально использовались балластные сопротивления в пределах от 100 кОм до 5.1 МОм, в то время как сопротивления образцов как минимум на порядок величины больше).

При дальнейшем увеличении силы одноосного давления сопротивление образца становится сравнимым с балластным сопротивлением. Соответствующая осциллограмма приведена на Рис.3 .

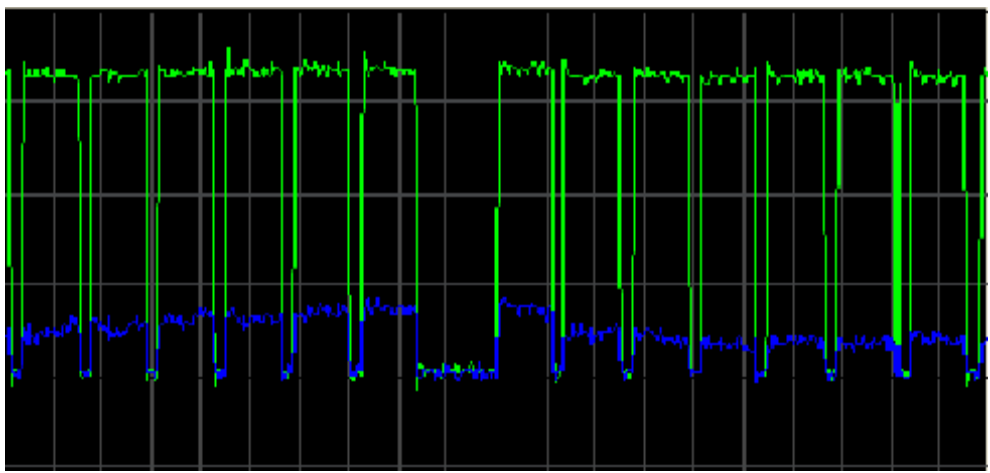

Рис.3 Осциллограмма двух каналов измерительной установки (см. Рис.1) для случая, когда сопротивление исследуемого образца пленки пластиката под действием одноосного давления становится сопоставимым с балластным сопротивлением.

В этом случае сигнал с полимерной пленки испытывает значительные нетермодинамические» флуктуации и может иметь медленные тренды к понижению, т.е. к уменьшению сопротивления образца при неизменном значении давления. Следует отметить, что в первом приближении описанный отклик образца при заданном значении давления слабо зависит от различных версий (в том числе и биполярной) импульсной последовательности. Соответствующая осциллограмма приведена на Рис. 4

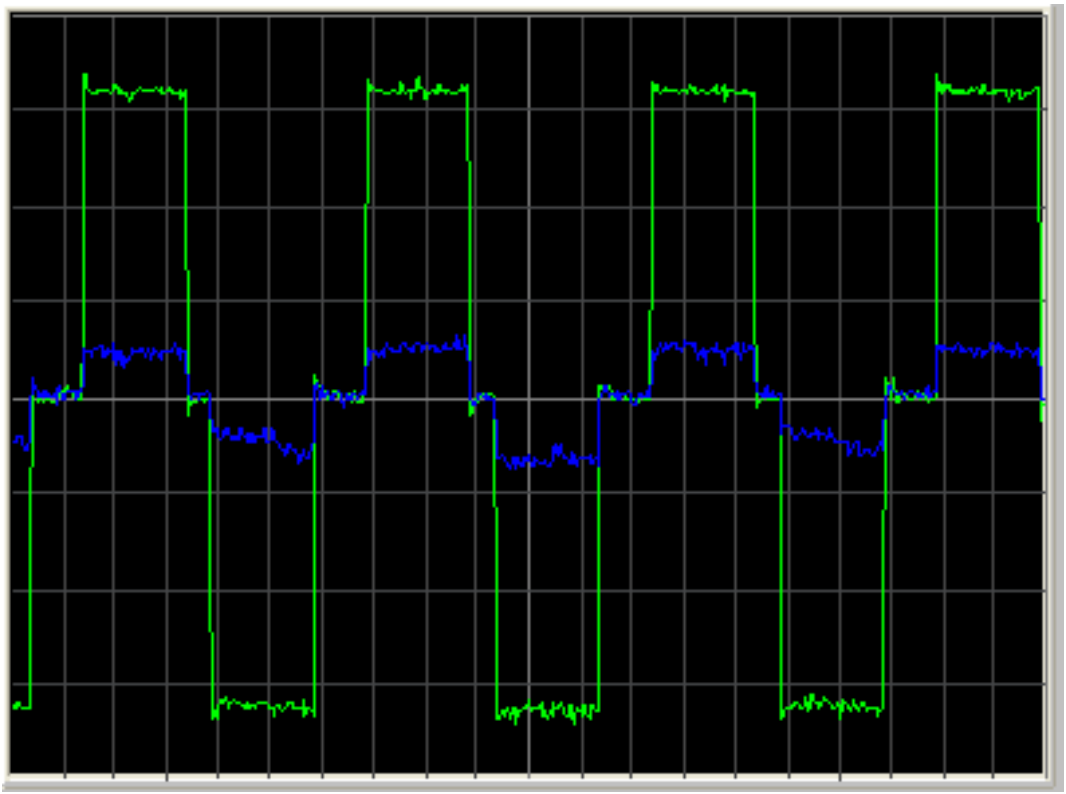

Рис.4 Реакция образца на двуполярную импульсную последовательность зондирующего напряжения. Кроме флуктуаций амплитуд импульсов заметны процессы релаксации образца или проявления процессов «установления».

В экспериментах с полимерными пленками различной толщины форма пробной импульсной последовательности на исследуемом образце мало отличалась от исходной. Среди общих закономерностей, установленных в настоящей работе, можно отметить, что для относительно тонких пленок порядка 30-50 мкм переход в СВП происходил при низких значениях одноосного давления, обратимо и практически скачком. Характерные значения силы давления, обеспечивающих переход в СВП для тонких пленок от 1 до 10 Ньютон. Практически только для таких пленок при попытках плавно и медленно отследить переход в СВП наблюдались неустойчивости и переходные процессы, в рамках которых напряжение на образце существенно отклонялось от программно задаваемой последовательности.

Так, на Рис.5 можно отчетливо наблюдать возбуждение генерации в момент, когда сопротивление образца заметно уменьшается под действием одноосного давления и становится сопоставимым с сопротивлением балласта. Генерация колебаний напряжения на образце срывается на фронтах пробного импульса.

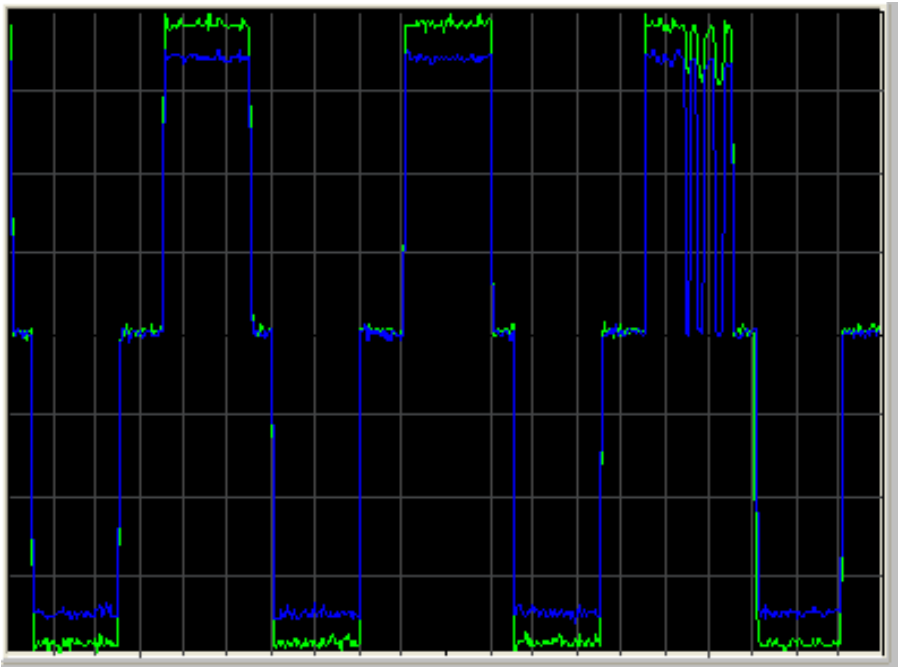

Рис.5 Возбуждение генерации в процессе перехода образца в СВП.

При дальнейшем увеличении силы давления происходит устойчивое переключение образца пленки пластиката в СВП, его сопротивление при этом становится существенно меньше балласта и напряжение на образце перестает отслеживать исходную последовательность, поскольку все падение напряжения происходит на балластном сопротивлении, а сопротивление образца приближается к аппаратному нулевому значению. Характерная осциллограмма перехода образца в СВП с сопротивлением порядка десятков Ом приведена на Рис.6.

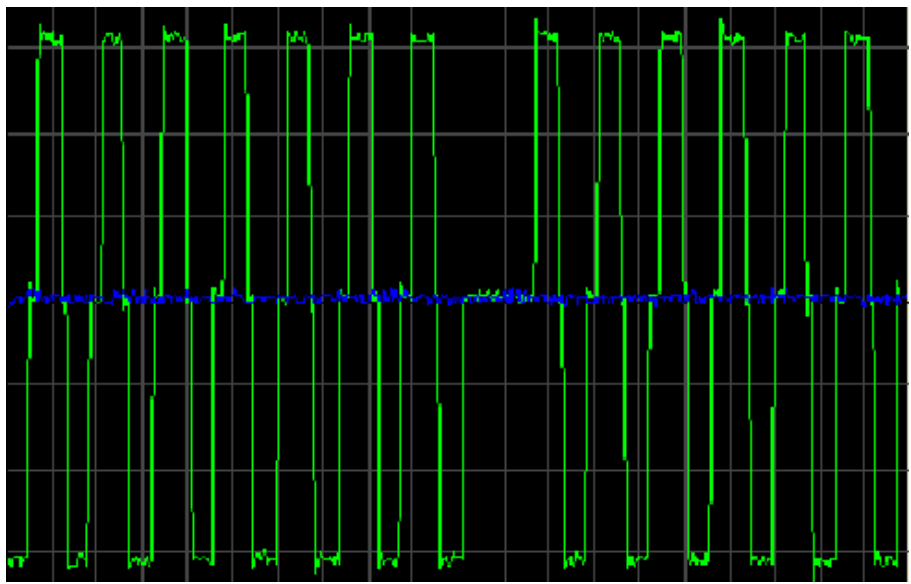

Рис.6 Осциллограмма, соответствующая переходу образца в СВП с сопротивлением на несколько порядков меньшим балластного сопротивления, равного в данном случае 100 КОм.

Приведенные выше экспериментальные осциллограммы использовались для качественного анализа временной структуры процессов, в то время как параллельно все данные осциллограмм в цифровом виде поступали в компьютер и после соответствующего усреднения и обработки использовались для получения количественных данных по изменению сопротивления образца в зависимости от величины одноосного давления. Характерная зависимость модуля тока, протекающего через образец при подаче импульсной последовательности амплитудой 10В, приведена на Рис. 7. Зависимость на Рис 7 получена для образца «средней» толщины порядка 250 мкм (соотношение ПВХ 100 : пластификатор «А» 80 по массе). Для полученных зависимостей тока для различных образцов, как правило, имеется гистерезис, хорошо воспроизводимый в последовательных экспериментах. При дальнейшем увеличении давления ток практически не изменяется и выходит на насыщение, причем сопротивление образца в СВП достаточно точно измеряется и оказывается несопоставимо большим сопротивления подводящих проводов, в отличие от результатов, известных для других широкозонных полимеров [6] .

Что касается более тонких пленок порядка 30-50 мкм, то переключение в СВП происходит при значительно меньших давлениях, величина гистерезиса меньше, но перепад сопротивлений образца до и после приложения давления значительно больше чем для более толстых пленок и достигает значений $10^4$ и более.

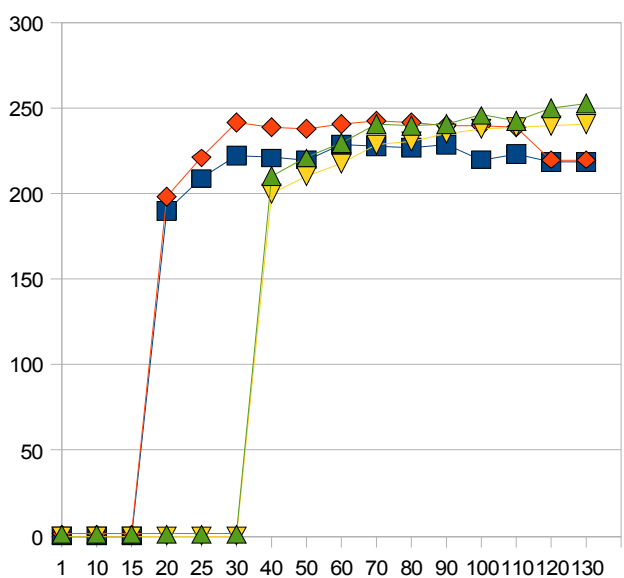

Рис.7. Характерная зависимость тока через образец пленки пластиката толщиной 250 мкм при приложении к нему одномерного давления. По оси ординат — сила давления F в единицах 10 Ньютон, по оси абсцисс ток через образец в произвольных единицах.

Напротив, при увеличении толщины пленки до 500 мкм зависимость от давления становится более плавной, гистерезис практически исчезает, а перепад сопротивлений до и после перехода в СВП сокращается. В настоящей работе не было специальной задачи найти «критические» толщины пленок пластикатов, для которых эффект перехода в СВП полностью отсутствовал. Следует отметить, что для антистатических пластифицированных пленок ПВХ, в нашем диапазоне напряжений и давлений зарегистрировать небольшие изменения тока при приложении одноосного напряжения удавалось для пленок толщиной до 1000 мкм.

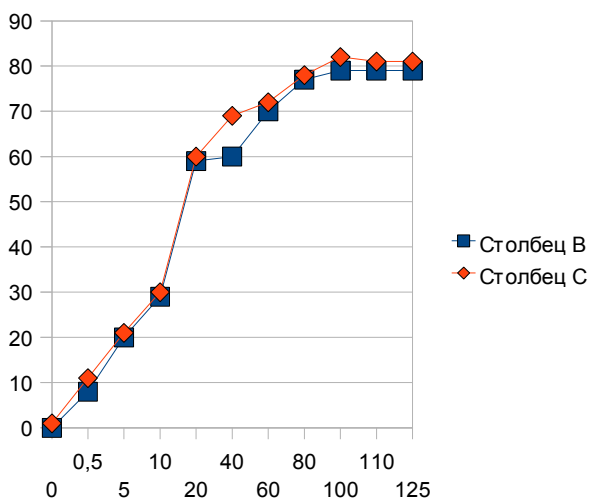

Рис.8. Зависимость тока через «толстый» образец порядка 500 мкм. Скачок проводимости практически исчезает, и возникает почти пропорциональная зависимость тока от прикладываемого одноосного давления.

Отметим, что в зависимости от толщины пленки и концентрации пластификатора, конкретные значения перехода в СВП при приложении одноосного давления могут существенно изменятся: так, для стандартного пластификатора диоктил-фталата (ДОФ) реализовать переключение в СВП пленок толщиной 30 мкм в СВП в используемом диапазоне давлений до $F= 1.5*10^3$ Ньютон и напряжениях импульсной последовательности 10В, не удалось. Аналогичные пленки толщиной 30 мкм, полученные с широко известным пластификатором дифенил-п–третбутил-фенилфосфатом (ДФИБФФ), продемонстрировали переход в СВП при значении силы $F=7*10^2$ Ньютон с диаграммой тока аналогичной приведенной на Рис.7

## 3. Обсуждение результатов

В приведенных выше экспериментальных результатах на качественном и количественном уровнях показано, что для пленок пластикатов ПВХ с антистатическим уровнем проводимости отклик на одноосное давление наблюдался для всех испытуемых образцов вплоть до толщин порядка 1 мм. Дальнейшее увеличении толщины пленки образцов затруднялось тем, что на поверхности пленок, получаемых методом полива из раствора, при таких толщинах возникали более или менее явные плавные неоднородности, и модель плоской пленки, в каждой точке поверхности соприкасающейся с электродом, переставала соответствовать условиям эксперимента. Качественный анализ всех последовательных стадий переходов пленки пластиката в СВП и обнаружение неустойчивых режимов генерации колебаний, на наш взгляд, полностью исключает возможность артефактов (например, замыкания шероховатостей контактов) в описанных выше исследованиях. Маловероятными представляются также трактовки перехода в

СВП, связанные с ростом металлических усов или образованием сквозных микроканалов и микротрещин в пленках пластикатов, что обусловлено как значительной толщиной образцов, так и их достаточно высокой пластичностью. Таким образом, на наш взгляд, объяснение наблюдаемого перехода полимерных пленок в СВП под действием одноосного давления, следует искать в физико-химических и структурных особенностях строения полимерных пленок

## 4. Электронно-молекулярная модель электропроводности ПВХ пластикатов

Как уже отмечалось, общая теория наблюдаемых выше и аналогичных явлений в полимерных пленках в настоящее время отсутствует. Для объяснения на качественном уровне перехода в СВП в гомогенных пластифицированных ПВХ пленках можно предложить следующую физическую модель.

Как известно, пластифицированный ПВХ имеет сложную микро и макро-молекулярную структуру, которая не может быть отнесена ни к упорядоченным, ни к полностью хаотичным. Пленки пластиката, полученные из истинного раствора полимера с пластификатором, являются неравновесными структурами, представляют собой аморфные полимеры, которые можно рассматривать как фрактальные структуры [19,20]. Известно стремление конденсированных макромолекулярных систем к самоорганизации в масштабно-инвариантных мульти-фрактальных формах [20]. Размеры образующихся клубков макромолекул определяются топологическими ограничениями движения молекул, «зацепами» и заматыванием пластифицированных, т.е. гибких макромолекул вследствие теплового движения и могут иметь размеры порядка нескольких характерных длин подвижного сегмента макромолекулы. В работе [20] сделана достаточно убедительная попытка показать, что формирующаяся в полимере квазирешетка клубковых супрамолекулярных образований определяет, в частности, объемную сжимаемость и сдвиговую жесткость полученного из раствора образца полимерной пленки.

Таким образом, макроскопически однородный пластикат на надмолекулярном уровне можно рассматривать как сложную и сильно неоднородную неравновесную многофазную (как минимум двухфазную) фрактальную систему, в которой можно выделить слабоупорядоченную за счет топологических ограничений квазирешетку более плотных клубковых супрамолекулярных образований, в которых свободное перемещение молекул и их сегментов строго ограничено, и аморфную фазу, содержащую свободные гибкие макромолекулы полимера и пластификатора. Для завершения построения модели пленки пластиката остается сделать достаточно разумное предположение, что сильные флуктуации плотности и большое число зацеплений в супрамолекулярных кластерах создают условия для развития процесса делокализации электронов, ограниченной размерами кластера, т.е. появления локально квазиметаллической проводимости. В обоснование сделанного предположения можно использовать тот факт, что фактически использованная в работе [8] стабильность экситонов в электрическом поле можно трактовать как существование локальных проводящих областей, которые при наложении внешнего электрического поля порождают стабильное разделение зарядов.

Таким образом, мы приходим к модели неоднородной двухфазной среды, в которой проводящие кластеры (клубки или домены) перемежаются с аморфной фазой, проводимость которой существенно ниже. Далее, при наложении внешнего поля на краях изолирующих областей возникают поверхностные скопления зарядов противоположных знаков и образуются «микроконденсаторы», причем за счет описанного выше эффекта сильных неоднородностей поле внутри «микроконденсатора» может существенно превос-ходить внешнее приложенное поле. Ранее основная часть изложенной модели предлагалась авторами в работах [1-5].

Для объяснения в рамках предлагаемой модели монотонного увеличения проводимости пластиката при увеличении концентрации пластификатора, т.е. при увеличении количества подвижных сегментов макромолекул, добавим к существующей модели понятие «динамических ловушек». Суть этого понятия в том, что инжектированные в полимер из металлических электродов носители заряда, в качестве которых могут фигурировать как электроны, так и дырки, относительно свободно перемещаются в проводящих доменах и скапливаются вблизи достаточно размытых границ кластеров, поскольку изотропная фаза не обладают собственной проводимостью. Тем не менее, поскольку функциональность кластера (т.е. число торчащих из кластера «хвостов» макромолекул [20]) может быть велика, при взаимодействии с молекулами аморфной фазы «хвосты» могут передать заряд на «динамическую» ловушку, связанную с подвижным сегментом. При этом сам переданный заряд будет, очевидно, иметь нулевую подвижность относительно макромолекулы, но целиком сегмент- носитель заряда, вследствие присутствия молекул пластификатора, сможет легко под действием поля «микроконденсатора» перемещаться к его противоположно заряженной обкладке (см Рис.9). Далее соприкасаясь с хвостами макромолекул второго кластера, «динамическая» ловушка передает ему свой заряд.

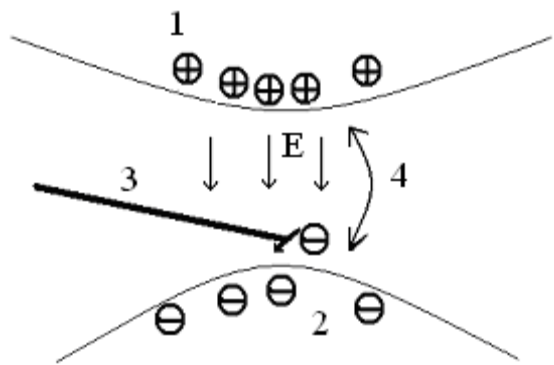

Рис.9. Схема переноса зарядов между проводящими кластерами 1,2 «динамической» ловушкой, связанной с подвижным сегментом макромолекулы 3. Стрелка 4 схематично показаны направления перемещения «динамической» ловушки.

Далее можно предположить, что вследствие броуновского движения сегмент с разряженной ловушкой вернется в исходное положение и перенос может повториться. Этот механизм переноса заряда через изолирующую аморфную среду является альтернативным по отношению к прямому туннелированию электрона через эффективный барьер, а кроме того, он уравнивает шансы переноса электронов и дырок и может объяснить наличие антистатического уровня проводимости в пленках пластикатов с модификатором «А», а также известный эффект увеличения проводимости (на уровне антистатики) с увеличением концентрации пластификатора.

При увеличении внешнего поля концентрация зарядов на обкладках «микроконденсатора» возрастает. Это приводит к увеличению вероятности того, что сегмент макромолекулы, приносящий заряд к обкладке «микроконденсатора», не только передает заряд под действием сил Кулона, но и может сам «подхватить» (возможно в другую ловушку) заряд противоположного знака. При этом возникает качественно новый режим переноса заряда, когда подвижный сегмент макромолекулы начнет двигаться (по стрелке 4 см. рис.9) под действием силы Кулона в обоих направлениях между обкладками, что и может приводить к значительному скачку проводимости в пленках пластифицированного ПВХ. Фактически в этом режиме при перезарядке динамической ловушки из отрицательного заряда в положительный на втором кластере возникают два электрона, т.е. процесс становится лавинообразным, и может возникать «мягкий пробой», т.е. скачкообразный переход в СВП. В принципе, роль «динамических» ловушек могут выполнять как поляризованные фрагменты, так и дефекты макромолекул.

В рассматриваемой модели обнаруженные в различных экспериментах [6-8] каналы проводимости могут образовываться посредством описанных выше подвижных ловушек, способных осуществлять перенос заряда между проводящими кластерами, причем последовательное развитие таких каналов может приводить в конечном итоге к образованию более длинного канала, замыкающегося на металлические электроды.

Пластичность и подвижность доменов в рамках квазирешетки и изолирующих промежутков, связанная с подвижностью сегментов молекул полимера и обусловленная наличием пластификатора, позволяет предположить медленные изменения конфигурации цепочки «микроконденсаторов», что при уменьшении ширины «максимального» изолирующего слоя может приводить к спонтанным переходам канала как в СВП, так и наоборот. Низкая подвижность границ кластеров и взаимное притяжение поверхностных зарядов противоположных знаков на обкладках «микроконденсаторов» могут объяснить сохранение состояния СВП при снятии внешнего поля, т.е. собственно эффект памяти.

Для объяснения эффекта перехода в СВП при воздействии одноосного давления, отметим, что в полимере даже при слабом трении происходит эффективное разделение зарядов (например в быту типичные статические напряженности электрического поля могут достигать киловольт). Собственно из-за этой способности полимеров к накоплению свободных зарядов антистатики и получили широкое распространение. Таким образом, достаточно учесть, что внешнему давлению будет оказывать сопротивление в основном квазирешетка макромолекулярных кластеров, которая может создавать противодавление за счет кулоновских сил, т.е. скоплений зарядов на границах кластера. Благодаря этому появляются свободные заряды на обкладках «микроконденсаторов», что аналогично приложению внешнего поля. При достаточно интенсивном давлении разделение зарядов в среде достигает величины, когда существенную роль могут играть описанные выше «динамические ловушки», и при приложении внешнего пробного поля полимерный образец демонстрирует высокую проводимость. Очевидно, что чем больше толщина пленки, тем дольше и труднее будет выстраиваться цепочка комбинаций кластеров-ловушек и тем выше будет результирующее сопротивление полимерного образца, что и наблюдалось в описанных выше экспериментах.

5. Заключение

Показано, что скачкообразное увеличение проводимости в пленках пластикатов ПВХ реализуется в широком интервале толщин образцов, причем при увеличении толщины пленки пороговое одноосное давление, соответствующее скачку проводимости также возрастает. Полученные для пластикатов ПВХ данные существенным образом отличаются от достаточно многочисленных публикаций, где в качестве критической толщины (т.е. при использовании более толстых пленок эффект отсутствует) указывается значение 1-3 мкм, что способствует получению более полной информации об аномалиях электропроводности в наблюдаемых системах металл-полимер-металл.

Полученные результаты позволили предложить качественную модель наблюдаемых аномалий проводимости, объясняющую на качественном уровне практически все полученные в экспериментах авторов результаты. Весьма вероятно, что фрагменты предложенной модели могут оказаться полезными для объяснения аномалий электропроводности и для других классов широкозонных полимеров.

Предложенная выше модель прыжковой проводимости с участием «динамических ловушек» позволяет качественно объяснить все основные особенности переходов в СВП, наблюдаемые в пленках пластифицированного ПВХ, и может найти применение и в случае других аналогичных полимеров. Тем не менее, как и другие модели проводимости, предложенные в литературе, данная модель нуждается в дополнительной экспериментальной проверке и требует развития для получения количественных оценок.